\documentclass[twocolumn,aps,showpacs,letterpaper]{revtex4}
\usepackage{graphicx}

\begin{document}

\hyphenation{Ka-pi-tul-nik}

\title{New Experimental Constraints on~Non-Newtonian Forces below $100$~$\mu$m}
\author{J.~Chiaverini, S.~J.~Smullin, A.~A.~Geraci, D.~M.~Weld, and  A.~Kapitulnik}
\affiliation{Departments of Applied Physics and of Physics,
Stanford University, Stanford, CA 94305}

\date{\today}

\begin{abstract}

We have searched for large deviations from Newtonian gravity by
means of a finite-frequency microcantilever-based experiment. Our
data eliminate from consideration mechanisms of deviation that
posit strengths $\sim\nobreak10^4$ times Newtonian gravity at
length scales of $20$~$\mu$m. This measurement is 3 orders of
magnitude more sensitive than others that provide constraints at
similar length scales.

\end{abstract}

 \maketitle

Physics beyond the standard model will be explored at the Large
Hadron Collider in a few years, but gravitational measurements at
small length scales are capable of investigating this region now.
These measurements become particularly important in light of
recent theoretical developments.  Moduli, massive scalar particles
that could mediate forces with strengths possibly $\sim 10^6$
times Newtonian gravity, could have Compton wavelengths in the
range of $10$~$\mu$m--$1$~mm. These particles would manifest
themselves as strong gravitylike forces. The hierarchy problem,
the seeming disparity ($16$~orders of magnitude) between the
standard model energy scale and the Planck scale, suggests other
reasons to study this regime. This problem can be recast by asking
why gravity is so weak compared to the other known forces in
nature. Recent theoretical work~\cite{add} suggests that extra
spatial dimensions, possibly as large as $1$~mm and accessible
only to gravitons, may supply a solution. Because of spreading
into the extra dimensions, gravity would be diluted compared to
the other forces, and therefore would seem quite weak. As gravity
has not been well tested below the centimeter scale until
recently~\cite{adel,review}, there are many reasons to investigate
gravitylike forces at length scales of $\lesssim 100$~$\mu$m.

Exotic massive scalar particles would modify the standard
Newtonian potential with an additional Yukawa-type
term~\cite{scalar}. A similar correction would exist for
observation of gravity at length scales roughly the size of any
compactified extra dimensions~\cite{add}.  This leads to the
following equation for the gravitational potential in the presence
of such non-Newtonian effects:

\begin{equation}
\label{potential} V=- G {m_1 m_2 \over r}(1 + \alpha e^{-{r /
\lambda}})
\end{equation}

\noindent Here $G$ is Newton's constant, $m_1$ and $m_2$ are the
masses, $r$ is the center of mass separation, and $\alpha$ is the
strength (relative to Newtonian) of any new effect with a length
scale of $\lambda$.  Non-Newtonian effects are typically
parameterized in terms of $\alpha$~and $\lambda$.

In this Letter we present data obtained for attracting-mass
surface separations down to $25$~$\mu$m, enabling investigation of
interaction scales below $\sim\nobreak10$~$\mu$m.


The experiment is reminiscent of torsion pendulum gravity
measurements, though an ac drive force is employed, and the masses
are much smaller. In place of the typical torsional fiber force
sensor, a microcantilever was used. A mass attached to this
cantilever was subjected to a time-varying force of a
gravitational type, and the deflection of the cantilever (measured
using fiber-optic interferometry~\cite{rugar}) provided a measure
of this force. The experiment was performed at low temperature and
in cryogenic vacuum to exploit the high quality factors attainable
in microcantilevers under these conditions.

The force sensors for these experiments were single-crystal
silicon diving-board-shaped oscillators $250$~$\mu$m in length,
$50$~$\mu$m in width, and $0.335$~$\mu$m thick.  The cantilevers
were fabricated using standard micromachining
techniques~\cite{atto} and have quality factors in the $10^3-10^5$
range (in vacuum).  Spring constants of these sensors were
$5.0$--$5.5$~mN/m, and the resonant frequency of the lowest
flexural mode of the cantilever was shifted from $\sim 7000$~Hz to
$\sim 300$~Hz upon mass attachment.

Microcantilever sensitivity is usually limited by thermal noise,
i.e., Brownian motion of the beam.  This noise limit, analogous to
the Johnson noise in a resistor, can be used to obtain the minimum
detectable force of a cantilever: $F_{min}=\sqrt{ 4 k_B T k B /
\omega _0 Q }$. Here $k_B$ is the Boltzmann constant, $T$ is the
temperature, $k$ is the spring constant, $\omega _0$ is the
resonant frequency (rad/s), $Q$ is the quality factor, and $B$ is
the measurement bandwidth. The typical force sensitivity for these
(thermal noise limited) measurements was $\sim\nobreak1 \times
10^{-16}$~N/$\sqrt{\rm Hz}$. With averaging times of a few hours,
the present devices could reach an ultimate sensitivity below
$\sim\nobreak1\nobreak\times 10^{-17}$~N ($200$~times the expected
Newtonian gravitational force).

\begin{figure}
\includegraphics[width=0.95 \columnwidth]{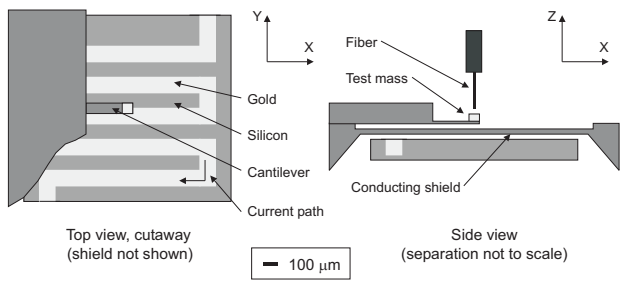}
\caption{Schematic top and side views of test mass on cantilever
and drive mass below, showing alternating gold and silicon drive
mass bars.  Drive mass motion was in the $Y$ direction.}
\label{schemmag}
\end{figure}

The force measured was that between a mass placed on the end of
the cantilever (the test mass) and a larger mass oscillated a
small distance away (the drive mass).  The test masses were gold
rectangular prisms, $50\nobreak\times 50 \times 30$~$\mu$m$^3$ in
size and $\sim\nobreak1.4$~$\mu$g in mass, and were attached to
the cantilevers with a thin (typically $\lesssim\nobreak1$~$\mu$m)
layer of epoxy.  The drive mass was constructed of ten bars: five
gold bars alternating with five silicon bars. Each bar was
$100$~$\mu \rm m\times 100$~$\mu \rm m \times 1$~mm long. (See
Fig.~\ref{schemmag} for schematic representation.) This
construction, when oscillated in the plane of the drive mass and
perpendicular to the direction of the long dimension of the bars,
is expected to gravitationally excite the test mass at a harmonic
of the oscillation frequency.  This shift of the gravitational
signal frequency with respect to that of the drive mass
oscillation helped prevent spurious excitation of the cantilever
due to unwanted vibration at the signal frequency.

To test the system and precisely align the drive mass with the
test mass on the cantilever, as well as to provide an {\it in
situ} equivalent measurement, a magnetic analog of the
gravitational experiment was used.  The gold bars in the drive
mass were connected electrically at alternating ends (see
Fig.~\ref{schemmag}) to form a meander. When a dc electric current
is driven through the path defined by the gold bars and their
interconnects, a space-varying magnetic field (with a spatial
periodicity half that of the gravitational response) is created
above the drive mass.  This field couples to the test mass through
an evaporated layer of nickel on the test-mass surface closest to
the drive mass.  The oscillation of the drive mass creates a
time-varying magnetic field at the location of this layer. A
signal proportional to current through the meander verifies the
system's functionality.

To fabricate drive and test masses, bulk silicon was etched to
form a mold into which gold was evaporated. After evaporation, the
mold was ground and polished. In the case of the test masses, the
silicon was then dissolved to release the gold prisms (for
fabrication details, see~\cite{our_rsi}).

The oscillation of the drive mass beneath the test mass was
facilitated by attachment of the drive mass to the free end of a
clamped piezoelectric bimorph actuator. The actuator was $1$~mm
thick and $4$~mm wide and had a free length of $41$~mm.

The actuator's motion was characterized {\it ex situ} using a
laser beam-bounce method.  In the experimental probe, the
actuator's motion was measured capacitively using a calibration
obtained from this initial characterization.  The actuator's drive
frequency was tuned to one-third of the cantilever resonant
frequency by means of magnetic excitation. The amplitude of
actuator motion was 98$\pm$7~$\mu$m.  Finite element calculations
were performed to determine the expected Newtonian and
non-Newtonian response of the test mass.


The nonlinearity in the piezoelectric actuator caused a small
amount ($4\%$) of actuator vibration at the cantilever resonance
(the third harmonic of the actuator frequency) and could lead to a
spurious signal.  Therefore, two mass-spring vibration isolation
stages separated the cantilever mount from the actuator mount.
Each stage had a resonant frequency of $\sim\nobreak2$~Hz for both
vertical and horizontal motion, and thus together the stages
provided an attenuation of $\sim\nobreak10^8$ at $300$~Hz between
the mounts.

A stiff metallized shield was placed between the drive and test
masses to minimize the effects of electrostatic and Casimir forces
by preventing direct ac coupling between the masses. The shield
was a  $2.8$~mm by $5.2$~mm silicon-nitride membrane $3$~$\mu$m
thick onto both sides of which was evaporated $100$~nm of gold.
Supported by a 1~cm$^2$ silicon wafer die, the shield was attached
to the cantilever wafer $15$~$\mu$m from the cantilever.


The vertical $(Z)$ and in-plane $(X, Y)$ relative displacements of
the test and drive masses were determined by means of a capacitive
position sensor (CPS) similar to that described
in~\cite{field_cps}. Two simpler capacitors, in conjunction with
the CPS, provided measurement of relative tilt. After cooling, the
position was adjusted until the original capacitive readings were
regained, using a three-axis translation stage. Because of the
drive mass geometry, the position in $X$ (the direction along the
drive mass bars) need only be determined to an accuracy of
$\sim\nobreak100$~$\mu$m. In the $Y$ direction, greater precision
was required as the gravitational response was expected to vary by
$\sim\nobreak100$\% over 50~$\mu$m in~$Y$. The CPS enabled
determination of relative position in this direction to
$\pm\nobreak2$~$\mu$m. The tilt was adjusted by differential
heating of the upper vibration isolation stage support springs.

After lateral realignment, an absolute value of the vertical
separation between the masses was determined by means of direct
mechanical contact (sensed by the cantilever) between the drive
mass and the shield. Subsequent motion away from the shield was
measured with the CPS, and the separation between the masses was
thus known to an accuracy of $2$~$\mu$m.

The cantilever, actuator, and vibration isolation stages were
suspended in a vacuum can at the end of a probe that was inserted
into a liquid helium research dewar. An exchange gas space
separated the inner vacuum can from the liquid helium in order to
reduce external vibrations due to helium boiloff.  The entire
system was supported by $2$~m long $\sim\nobreak2$~Hz springs for
additional isolation. The experiments were performed at
$T=9$--$11$~K, and at gas pressures of less than $10^{-4}$~torr
(room temperature reading) with further cryopumping while cold.

The sinusoidal actuator voltage and the cantilever displacement
signal were sampled and stored simultaneously at $10$~kHz using an
analog-to-digital converter. The displacement data stream was then
sorted into bins using the phase of the actuator drive voltage and
averaged by actuator phase. As the actuator phase is a measure of
the lateral position of the drive mass, this technique is similar
to a lock-in technique in that signals that are not phase coherent
with the actuator drive signal are rejected.

\begin{figure}
\includegraphics[width=0.9 \columnwidth]{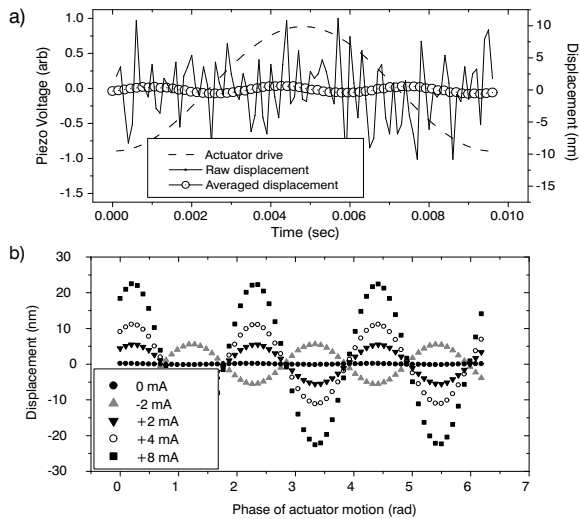}
\caption{(a) Actuator voltage (left axis), raw cantilever
displacement (right axis), and averaged cantilever displacement
(right axis, $10$~min averaging time) as a function of time over
one period of actuator motion. (b) Averaged data for one period of
actuator motion for five values of dc current through the drive
mass meander. } \label{rawdata}
\end{figure}

The third harmonic of the Fourier transform of the averaged data
is the motion of the cantilever that is phase correlated with the
third harmonic of the drive mass motion, i.e., the
gravitational-like response of the test mass \lbrack see
Fig.~\ref{rawdata}(a)\rbrack.  Using the spring constant and the
measured quality factor of the cantilever, the force on the test
mass due to the drive mass can be determined.

Data were collected as a function of drive mass meander current
and $Y$ position in order to align the drive mass to the test
mass. Figure~\ref{rawdata}(b) shows data for one period of the
drive mass actuator motion as a function of drive mass meander
current. The phase of the magnetic response changes by $\pi$ as
the current is changed from negative to positive. Also present is
a small asymmetry in the magnitude of the magnetic signal for
positive and negative values of the nominally same current
magnitude.  The observed zero-current signal adds phasewise to the
signal for a particular current, producing this asymmetry.

\begin{figure}
\includegraphics[width=0.9 \columnwidth]{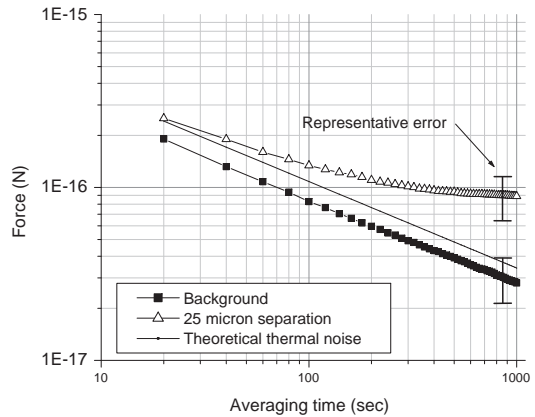}
\caption{Background and experimental signal, as well as
theoretical thermal noise, as a function of averaging time.  The
experimental signal levels off at a force of $8.9\times
10^{-17}$~N.} \label{datavtime}
\end{figure}

The third harmonic of two sets of data acquired over $1000$~sec
and averaged as a function of measurement time is plotted in
Fig.~\ref{datavtime} ($Q=1180$ for these data). Also plotted is a
line representing the theoretical thermal noise of the cantilever.
Data taken while the actuator was oscillating far from the test
mass match the theoretical curve within error~\cite{footnote1} and
imply that the dominant experimental noise is thermal in nature.
(Note that the theoretical curve depends on experimental
parameters.) Data taken with the drive mass oscillating and in
close proximity to the test mass show a signal that clearly is not
diminished with averaging and is patently above the thermal noise.
According to finite-element calculations yielding an expected
gravitational response of 4.0$\times10^{-20}$~N, this anomalous
signal is too large (by a factor of $\sim\nobreak2000$) to be of
Newtonian origin. Within experimental error, the magnitude of the
signal (relative to the actuator drive) as a function of
$Y$-direction offset is not consistent with a mass-dependent
force.

These data lead to the conclusion that the experiment's force
resolution is presently limited by an environmental effect, most
likely an electrostatic interaction between the shield and the
cantilever as discussed below. While the shield prevents direct
coupling of this sort between the masses, a secondary effect due
to shield motion could spuriously drive the cantilever. Other
possible non-mass-dependent effects are too small to appear at
this level. For instance, randomly quenched magnetic impurities in
the gold of the drive mass produce a moment that would result in a
force below the level of Newtonian gravity, and coordinated
pressure variations in the ballistic, residual gas in the probe
are similarly small.

Shield motion was characterized (using the fiber interferometer)
under standard experimental conditions. The response was not
sensitive to the current through the drive mass meander or
relative mass position. With the actuator swinging its full
amplitude, the shield motion at the first harmonic was
$\sim\nobreak10$~pm.  The signal at the third harmonic was an
order of magnitude less.


Motion of the membrane could drive the cantilever by creating an
ac Casimir force. An upper bound can be calculated assuming the
cantilever is a conductor and using the Casimir
result~\cite{casimirp} for two conducting planes (force per unit
area), $F_A\nobreak=-\pi ^2 \hbar c / 240 z^4$. For $10$~pm of
motion of the shield, the oscillatory force is on the order of
$1\nobreak\times 10^{-21}$~N, much less than the thermal noise of
the cantilever for experimental measurement times.

The motion of the membrane could also induce cantilever motion if
there were a significant potential difference between the
metallized top of the shield and the cantilever. Making a
worst-case assumption of two parallel plates with an area the size
of the cantilever, a voltage difference $0.3$~V would produce a
force on the cantilever of $1\nobreak\times 10^{-16}$~N.
Potentials of this size have been observed in other
experiments~\cite{stipe}, and we cannot yet exclude the
possibility that such a force currently limits our experimental
sensitivity. Shield motion was most likely due to inhomogeneities
in the drive mass surface caused by the use of electrically
dissimilar materials.  The design of upcoming experiments contains
insulating and conducting layers above the drive mass bars; the
conducting layer will shield the effect of any drive mass surface
variations.

\begin{figure}
\includegraphics[width=0.9 \columnwidth]{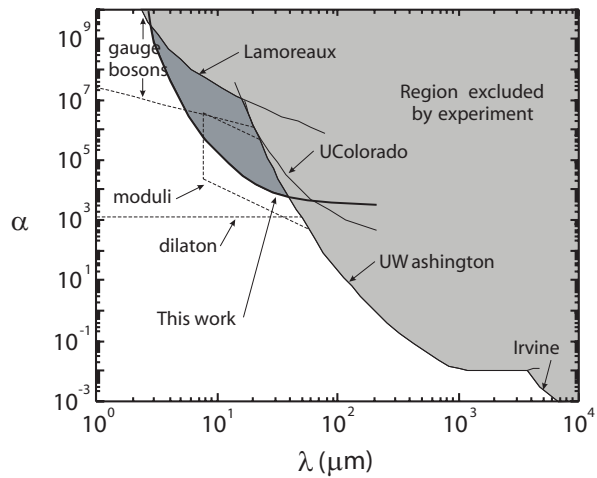}
\caption{ Strength versus length scale parameter space \lbrack see
Eq.~(\ref{potential})\rbrack\ for non-Newtonian effects showing
area excluded by present (darker shaded region bounded by signal
plus 2$\sigma$-level error) and previous (lighter shaded region)
experiments. Lines labeled Lamoreaux, UWashington, UColorado, and
Irvine are from~\cite{lamoreaux}, \cite{adel}, \cite{price},
and~\cite{irvine}, respectively. Theoretical predictions (dashed
lines) are adapted from~\cite{dilaton_ref,scalar}~(dilaton and
moduli) and from~\cite{add,savas}~(gauge bosons). }
\label{phasespace}
\end{figure}

The existence of a spurious force of average magnitude $8.9\times
10^{-17}$~N allows calculation of an upper limit on the magnitude
of any gravitylike force. Figure~\ref{phasespace} displays the
region of non-Newtonian interaction strength versus interaction
distance parameter space that can be excluded by these
results~\cite{footnote2}, as well as other current experimental
bounds.  For a length scale of $15$~$\mu$m, the present data
improve constraints on the strength of non-Newtonian effects by
$\sim\nobreak3$~orders of magnitude. The region eliminated is an
important one, as it contains previously unexplored space for
scalar moduli particles.


We have presented results from a search for gravitylike deviations
from Newtonian theory at distance scales below $100$~$\mu$m using
a microcantilever approach, with masses of size on the order of
these scales. Our data offer a new limit on non-Newtonian effects
in the range of $10$~$\mu$m, constraining theoretical predictions
of moduli.

We thank Thomas Kenny and Savas Dimopoulos for useful discussions.
J.C. and D.M.W. thank DoD for support. S.J.S. thanks NSF for
support. This work made use of the National Nanofabrication Users
Network facilities supported by the National Science Foundation
under Award No. ECS-9731294.

{\it Note added.}---After submission of this manuscript, results
for larger length scales were published from a similar experiment
at the University of Colorado \cite{pricenature}.


\begin{thebibliography}{xx}
\bibitem{footnote1}
The error displayed in Fig.~\ref{datavtime} is the sum of the
following errors in quadrature:  voltage measurement, 2\%; laser
wavelength, 0.1\%; interferometer fringe size, 5\%; quality
factor, 10\%; fiber position, 5\%; and spring constant, 10\%.
\bibitem{footnote2}
The error used in this determination is the sum of the following
errors in quadrature propagated through Eq.~(\ref{potential}) for
each value of $\lambda$: mass-to-mass distance, 4\%; actuator
amplitude, 7\%; gold density, 3\%; silicon density, 1\%; test and
drive mass lateral dimensions, 2\% and 1\%; test and drive mass
thickness, 5\% and 3\%.
\end{thebibliography}
\end{document}